%% file: main.tex
\documentclass[conference]{IEEEtran}
\IEEEoverridecommandlockouts
\usepackage{cite}
\usepackage{amsmath,amssymb,amsfonts}
\usepackage{algorithm}
\usepackage{graphicx}
\usepackage{textcomp}
\usepackage{xcolor}
\usepackage{balance}
\usepackage{multirow}
\usepackage{booktabs}
\usepackage{comment}
\usepackage{xspace}
\usepackage{algpseudocode}
\usepackage{subfig}
\usepackage{url}

\def\BibTeX{{\rm B\kern-.05em{\sc i\kern-.025em b}\kern-.08em
    T\kern-.1667em\lower.7ex\hbox{E}\kern-.125emX}}

\newcommand{\name}{CoCoDiff\xspace}

\begin{document}
\pagestyle{plain}

\title{\name: Optimizing Collective Communications for Distributed Diffusion Transformer Inference Under Ulysses Sequence Parallelism}

\author{
  \IEEEauthorblockN{Bin Ma\IEEEauthorrefmark{1}, Xingjian Ding\IEEEauthorrefmark{1}, Tekin Bicer\IEEEauthorrefmark{2}, Pengfei Su\IEEEauthorrefmark{1}, Dong Li\IEEEauthorrefmark{1}}
  \IEEEauthorblockA{\IEEEauthorrefmark{1}University of California, Merced, CA, USA \\
    \{bma6, xding4, psu7, dli35\}@ucmerced.edu}
  \IEEEauthorblockA{\IEEEauthorrefmark{2}Argonne National Laboratory, Lemont, IL, USA \\
    tbicer@anl.gov}
}


\maketitle
\thispagestyle{plain}

\begin{abstract}
Diffusion Transformers (DiTs) are increasingly adopted in scientific computing, yet growing model sizes and resolutions make distributed multi-GPU inference essential. Ulysses sequence parallelism scales DiT inference but introduces frequent all-to-all collectives that dominate latency. Overlapping these with computation is difficult due to tight data dependencies, large message volumes, and asymmetric interconnect bandwidths.

We introduce \name, a distributed DiT inference engine exploiting two observations: (1) $V$ requires only linear projection while $Q$/$K$ need additional normalization and RoPE, creating opportunities to overlap $V$'s communication with $Q$/$K$ computation; (2) adjacent denoising steps produce similar tensors, yielding temporal redundancy. \name introduces three mechanisms: Tile-Aware Parallel All-to-all (TAPA) decomposes collectives into topology-aligned phases; V-First scheduling hides $V$'s communication behind $Q$/$K$ computation; and V-Major selective communication transmits only active projections on slow interconnects. On the Aurora supercomputer with four DiT models across 1--8 nodes (up to 96 Intel GPU tiles), \name achieves an average speedup of 3.6$\times$, peaking at 8.4$\times$.

\end{abstract}


\input{text/intro}
\input{text/background}
\input{text/method}
\input{text/evaluation}

\input{text/related_work}

\input{text/conclusions}

\balance
\bibliographystyle{IEEEtran}
\bibliography{IEEEabrv, bin, li2}

\end{document}

%% file: text/intro.tex
\section{Introduction}
\label{sec:introduction}

Diffusion Transformers (DiTs) have reshaped modern generative AI. Models such as Stable Diffusion 3 (SD3)~\cite{esser2024sd3}, FLUX~\cite{flux2024}, Qwen-Image~\cite{wu2025qwenimagetechnicalreport}, and Sora~\cite{videoworldsimulators2024} deliver unprecedented image and video quality, powering creative tools used by millions. DiTs are also gaining traction in scientific computing: AERIS~\cite{aeris2025}, for example, an 80B‑parameter Swin DiT, has been deployed on the Aurora supercomputer for high-fidelity global weather and subseasonal-to-seasonal (S2S) forecasting.
The computational cost of DiT inference is driven by its patch‑token representation: the token count grows quadratically with image resolution, and each generation task requires 
tens of sequential denoising steps. Distributed inference across multiple GPUs therefore becomes essential to meet the resulting computation and memory demands.

\begin{figure}[t]
\centering
\includegraphics[width=\columnwidth]{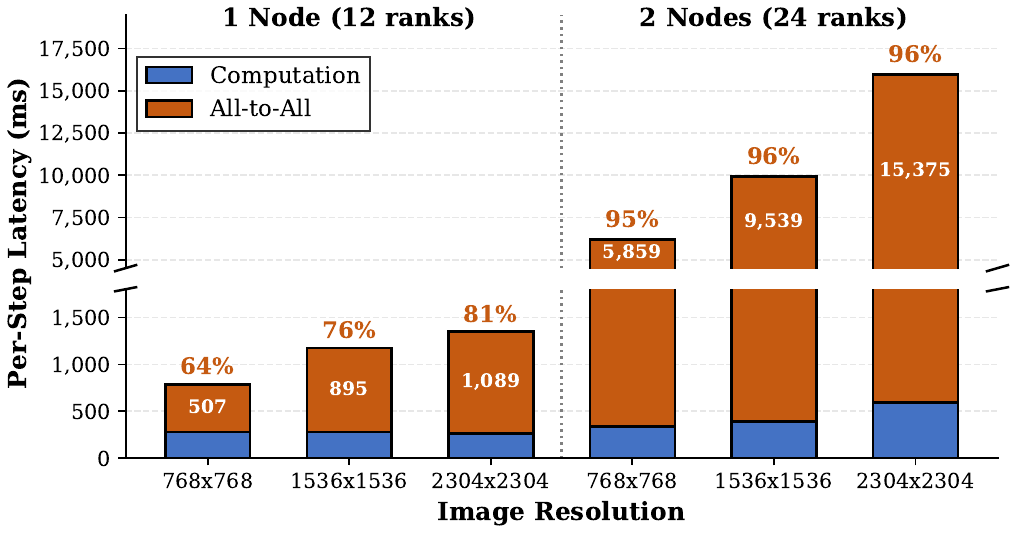}
\caption{Communication overhead in Qwen-Image (20B) inference on Aurora. The batch size is 1. The percentages indicate the percentage of all-to-all in overall inference time.}
\vspace{-15pt}
\label{fig:comm-overhead}
\end{figure}

Ulysses sequence parallelism~\cite{fang2024usp, liu2024ring} is the prevailing method for scaling DiT inference, sharding tokens across GPUs but requiring all‑to‑all exchanges of Q/K/V tensors before every attention layer, which is very expensive. Profiling Qwen‑Image 
(20B) on one node of the Aurora supercomputer (6 GPUs per node, each of which has 2 tiles) shows that all‑to‑all dominates performance: 64\% of inference time at the resolution $768\times768$ and over 80\% at the resolution $2304\times2304$ as shown in Figure~\ref{fig:comm-overhead}. 
This communication problem becomes even more pronounced with a higher resolution (i.e., a longer sequence), because the communication time complexity grows quadratically with the sequence length while computation grows only linearly (\S\ref{subsec:seq-parallel}). 
Multi‑node scaling further worsens the collective communication problem: with two nodes on Aurora, all‑to‑all reaches 96\% of inference time (shown in Figure~\ref{fig:comm-overhead}), as communication traffic traverses Aurora’s Slingshot fabric.

To reduce the overhead of collective communication,
overlapping communication with computation is an effective approach~\cite{ipdps19:jiawen}. However, this method faces challenges in the context of DiT. In essence, the computation graph and data dependencies in DiT are tightly coupled so that the collective communication cannot be easily overlapped with computation. 

First, in the sequence parallelism (especially the most popular Ulysses sequence parallelism), each transformer layer calls all-to-all to redistribute the tensors $Q$, $K$, $V$ across $P$ ranks before attention computation. This call blocks, even though it is a non-blocking version (e.g., \texttt{MPI\_Ialltoall}), until data from all ranks has arrived. 
Attention cannot start until the full redistribution finishes. 
This problem, in nature, comes from the dependency between attention and all-to-all \textit{within} a transformer layer. 


Second, there is dependency \textit{across} the layers, which limits the overlap opportunities. Diffusion transformers operate on sequential denoising steps, and each step requires the output of the previous one, which creates forward dependencies. Since the activations to a layer cannot be computed ahead of time, the communication sits directly on the critical path.  

Third, the communication is large and frequent, and does not decrease across denoising steps, which makes it difficult to amortize the communication costs. In particular, Ulysses-style parallelism communicates the tensors with the size $O(N \cdot d)$  at each layer and each denoising step where $N$ and $d$ are the sequence length and hidden dimension size respectively, which quickly saturates available bandwidth. 
The same full volume is repeated across all  denoising steps, even though adjacent steps produce  similar intermediate tensors.

The long-communication problem becomes even worse on the Aurora supercomputer where we did this work. Each node on Aurora is equipped with six Intel GPUs, each of which is equipped with two tiles. 
The inter-tile within the same GPU has high communication bandwidth (185 GB/s), 12$\times$ higher than inter-GPU communication bandwidth within the same node\cite{intel2023pvc, aurora2024alcf}. As a result, in a collective communication, the tiles not involved into the inter-GPU communication can finish the communication much earlier than other tiles, wasting tile-compute cycles. This problem comes from  the asymmetric communication topology across GPUs (or tiles on Aurora), and can be commonly found in other clusters where inter-node and intra-node have large difference in communication bandwidth.

To address the above challenges, we introduce \name, a high-performance, distributed inference engine for DiT. The design of \name is driven by two observations: (1) $V$ requires only linear projection while $Q$ and $K$, besides linear projection, additionally need normalization and RoPE, creating a time gap between the three tensors. (2) Adjacent denoising steps can produce highly similar intermediate tensors, showing strong temporal coherence, hence communications across adjacent steps can be redundant.


The first observation helps address the first and third challenges on the intra-layer dependency and communication volume. Instead of aiming to remove the dependency within a layer, we separate the collective communication for the three tensors ($Q$, $K$, and $V$) so that the collective communication for $V$ occurs first, right after the projection of $V$ is done. This early communication overlaps with the projection of $Q$ and $K$, hiding the communication overhead of $V$ and reducing the communication volume for the original \textit{lump} all-to-all for the three tensors.  

The second observation helps us address the second and third challenges on the inter-layer dependency and high communication frequency, and avoids negative impact of slow interconnects in asymmetric communication paths. In particular, we leverage the temporal redundancy across denoising steps such that those vectors in $Q$, $K$, and $V$ corresponding to similar tokens are not communicated, and are  reused across the denoising steps. This method leads to a substantial reduction of communication volume in slow interconnects, and has ignorable impact on inference accuracy. 

In conclusion, we summarize our contributions as follows.

\begin{itemize}
    \item We identify multiple challenges in collective communications that prevent scalable and high-performance distributed inference of DiT across GPUs. We introduce a high-performance inference engine, \name, to address those challenges.

    \item We leverage two salient characteristics of the DiT models to drive our solution with the considerations of interconnect topology and computation dependency.
    
    \item Evaluating on Aurora with four DiT models across 1--8 nodes (up to 96 Intel GPU tiles), \name achieves an average speedup of 3.6$\times$, peaking at $8.4\times$.
\end{itemize}

%% file: text/background.tex
\section{Background and Motivation}
\label{sec:background}


We review background and motivate our work. Table~\ref{tab:notation} summarizes the key notation used throughout this paper.

\begin{table}[!t]
\centering
\footnotesize
\setlength{\abovecaptionskip}{3pt}
\setlength{\belowcaptionskip}{3pt}
\setlength{\tabcolsep}{8pt}
\renewcommand{\arraystretch}{1.2}
\caption{Notations.}
\label{tab:notation}
\begin{tabular}{l@{\hskip 40pt}p{1.8in}}
\toprule
\textbf{Notation} & \textbf{Description} \\
\midrule
\multicolumn{2}{l}{\textbf{Attention \& Sequence Parallelism}} \\
$N$ & Sequence length (number of tokens). \\
$d$ & Per-head dimension size. \\
$H$ & Total number of attention heads. \\
$P$ & Number of tiles (ranks). \\
$h_i$ & Hidden state of token $i$. \\
$W_Q, W_K, W_V$ & Learned projection matrices $\in \mathbb{R}^{d \times d}$. \\
$q_i, k_i, v_i$ & Query, key, value vectors of token $i$. \\
$Q, K, V$ & Query, key, value matrices $\in \mathbb{R}^{N \times d}$. \\
\midrule
\multicolumn{2}{l}{\textbf{Denoising \& Selective Communication}} \\
$T$ & Total number of denoising steps. \\
$t$ & Current denoising step index. \\
$r$ & Cache ratio (fraction of projections reused). \\
$V_t^{p1}$ & $V$ after Phase~1 exchange at time step $t$. \\
$X_t^{p2}$ & $X \in \{Q,K,V\}$ after Phase~2 at time step $t$. \\
\bottomrule
\end{tabular}
\end{table}

\subsection{Ulysses Sequence Parallelism in Transformers}
\label{subsec:seq-parallel}

The attention mechanism lies at the heart of transformer architectures. Given an input sequence of $N$ tokens, the query matrix ($Q$), key matrix ($K$), and value matrix ($V$) (where $Q, K, V \in \mathbb{R}^{N \times d}$) are obtained via learned linear \textit{projections} (detailed in \S\ref{subsec:qkv-asymmetry}), where $d$ is the dimension size of an attention head. Attention then computes as follows:

\begin{equation}
\text{Attention}(Q, K, V) = \text{softmax}\left(\frac{QK^T}{\sqrt{d}}\right) \cdot V
\label{eq:attention}
\end{equation}

The result of Equation~\ref{eq:attention} leads to the attention matrix with the size of $N \times N$, yielding $O(N^2)$ memory complexity, which becomes prohibitive for long sequences.

Ulysses sequence parallelism addresses this by partitioning the input sequence along the sequence dimension across $P$  devices. Instead of each device processing the entire sequence length, each device processes a subset of tokens (i.e., $N/P$ tokens), allowing for longer sequence or larger model without out-of-memory errors on devices. We call this distributed data layout across devices, the \textit{sequence-parallel layout}. Ulysses~\cite{jacobs2023deepspeed}, widely employed in DeepSpeed/Megatron\cite{shoeybi2019megatron}, vLLM~\cite{Pagedatt}, and hugging Face Accelerate~\cite{accelerate2022}, is arguably the most popular sequence parallelism.
With Ulysses, before attention, an all-to-all collective communication transforms the sequence-parallel layout to the \textit{head-parallel layout}: each rank holds all $N$ tokens but only $H/P$ heads where $H$ is total number of heads in a layer. The attention proceeds on each rank's head subset. After the attention, a second all-to-all collective communication transforms from the head-parallel layout back to the sequence-parallel layout. 

\subsection{QKV Processing Asymmetry in Attention}
\label{subsec:qkv-asymmetry}

The attention mechanism requires three input tensors: $Q$, $K$, and $V$. They are generated through token projection. In particular, each token $i$ has a hidden state $h_i \in \mathbb{R}^{d}$, which is projected into a query vector $q_i = h_i W_Q$, a key vector $k_i = h_i W_K$, and a value vector $v_i = h_i W_V$, where $W_Q, W_K, W_V \in \mathbb{R}^{d \times d}$ are learned weight matrices. Stacking all $N$ tokens yields the matrices $Q, K, V \in \mathbb{R}^{N \times d}$.

\begin{figure}
\centering
\includegraphics[width=.93\columnwidth]{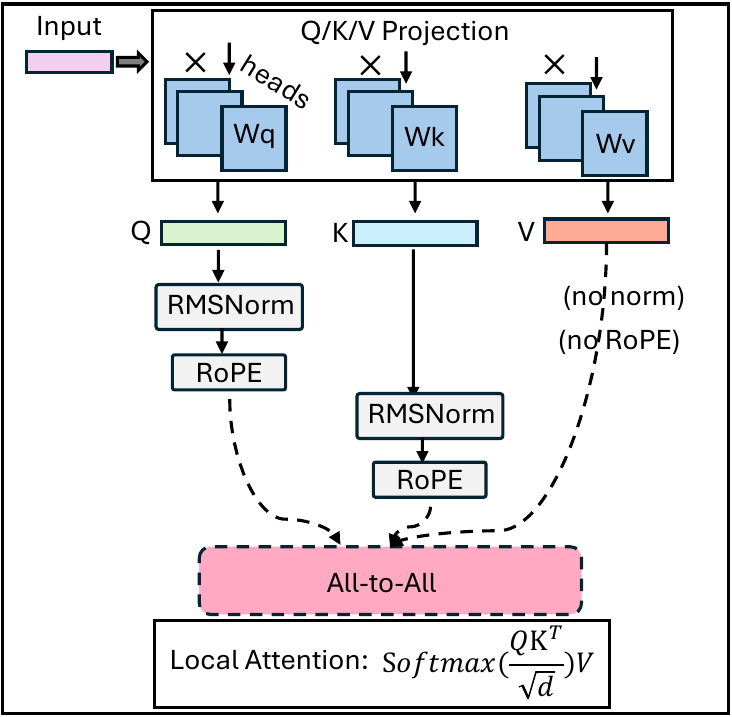}
\caption{QKV processing in the attention.}
\label{fig:vfirst-scheduling}
\vspace{-10pt}
\end{figure}

After the above projection, $Q$, $K$, and $V$ are reshaped to attention heads. Modern DiT architectures then apply additional processing \emph{exclusively} to $Q$ and $K$, creating computation asymmetry across the three tensors. 
The asymmetry comes from two operations for $Q$ and $K$, shown in Figure~\ref{fig:vfirst-scheduling}.

\textbf{QK normalization.} Modern DiTs, including SD3~\cite{esser2024sd3}, FLUX~\cite{flux2024}, and Qwen-Image~\cite{wu2025qwenimagetechnicalreport}, apply RMSNorm or LayerNorm to $Q$ and $K$ after the linear projection. The QK normalization controls the magnitude of $Q$ and $K$, so that the dot-product attention logits $QK^T$ remain numerically stable. $V$ does not participate in the normalization, because $V$ is built by weighted-summation of the resulting attention scores, and normalizing $V$ is meaningless.

\textbf{Rotary position embeddings (RoPE).} RoPE~\cite{su2024roformer} encodes positional information by rotating $Q$ and $K$ so that the dot product $q_i^T k_j$ depends on relative position $i - j$, where $q_i$ and $k_j$ are the query and key vectors at positions $i$ and $j$ respectively. Because the positional relationships are captured entirely through the product $QK^T$, applying RoPE to $V$ would alter the content being aggregated without contributing any positional signal. Hence, RoPE is not applied to $V$.

The above two operations are common, and they reflect fundamental design principles of modern transformers. Most of DiT have the two operations.

\subsection{Temporal Redundancy in Diffusion Denoising}
\label{subsec:temporal-redundancy}

Diffusion models generate images through iterative denoising over $T$ steps. Starting from Gaussian noise $x_T$, each step $t$ applies a learned denoising function to produce a progressively cleaner latent $x_{t-1}$. With DiT, this manifests as a sequence of attention computations whose intermediate tensors ($Q_t$, $K_t$, $V_t$) evolve gradually across steps. During this process, $Q$, $K$, and $V$ can show value similarity across two neighbor steps.  Hence, selectively reusing them across time steps can reduce the latency while causing minor impacts on the model accuracy. This is called the temporal redundancy, and has been leveraged to build a caching mechanism across step steps (e.g., DeepCache~\cite{ma2024deepcache}, $\Delta$-DiT~\cite{chen2024deltadit}, Learning-to-Cache~\cite{ma2024learningtocache}, and AttMEMO~\cite{feng2023attmemoacceleratingtransformers}).

The temporal redundancy has been exploited for compute savings on a single GPU, but not for communication reduction in distributed settings. Standard all-to-all implementations communicate all $Q$, $K$ and $V$ at full volume every step, regardless of how little individual vectors in $Q$, $K$ and $V$ have changed across consecutive steps. 

\subsection{Intel Aurora Architecture}
\label{subsec:aurora}

We use the Aurora supercomputer in our study. Each node in Aurora contains six Intel Data Center GPU Max 1550 (Ponte Vecchio) accelerators. Each GPU comprises two independent compute tiles with separate execution units and 64\,GB HBM2e memory, appearing as separate devices from software's perspective---yielding 12 addressable compute units per node. In our distributed inference setup, each tile is mapped to one rank, so a single node runs 12 ranks.

\begin{figure}[!t]
\centering
\includegraphics[width=\columnwidth]{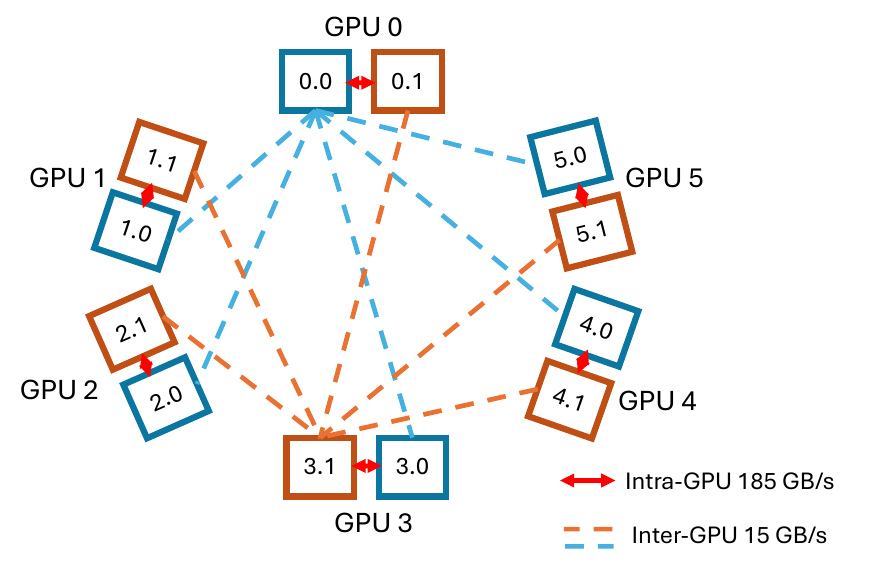}
\caption{The two-level communication hierarchy in an Aurora node. Intra-GPU bandwidth between tiles (red, solid arrows) is 12$\times$ higher than that of inter-GPU Xe Link (dashed lines).}
\vspace{-15pt}
\label{fig:aurora-topology}
\end{figure}

The communication fabric in Aurora exhibits a hierarchical structure. Within a single GPU, two tiles (Tile 0 and Tile 1) communicate through a high-bandwidth internal interconnect achieving unidirectional bandwidth of 185\,GB/s (or 269\,GB/s for bidirectional). Between GPUs, Intel's Xe Links connects all six GPUs in a mesh topology at unidirectional bandwidth of 15\,GB/s (or 23\,GB/s for bidirectional) per link. The Xe Links in a node form two independent rings: one connecting all Tile~0 in each GPU, and the other connecting all Tile~1 in each GPU, but there is no direct link across the two rings. 
For cross-node communications, each node has 8 HPE Slingshot-11 NICs providing unidirectional bandwidth of 25\,GB/s per NIC (or 200\,GB/s aggregated per node). The resulting three-level hierarchy (185 : 15 : 25\,GB/s unidirection) exhibits up to  12$\times$ difference in bandwidth across the hierarchy. 
Figure~\ref{fig:aurora-topology} illustrates this hierarchy. 

\subsection{Limitations of Existing Collective Communication}
\label{subsec:limitations}

Some implementations of collective communications consider interconnect topologies to improve performance. For example, NCCL~\cite{nccl2025} considers the communication hierarchy by using tree- or ring-based algorithms~\cite{thakur2005optimization,cho2019blueconnect}. However, those implementations optimize data movement without considering data access pattern and data semantics in computation, and no existing library fully fits Aurora's hierarchy. For example,  NCCL~\cite{nccl2025} does not offer a dedicated all-to-all communication primitive. The user of NCCL to build all-to-all has to use point-to-point communications,  \texttt{ncclSend} and \texttt{ncclRecv},  without hierarchical staging. MPICH~\cite{mpich} uses flat algorithms (Bruck, pairwise, and scattered), and its GPU direct-transfer treats all tiles as equal peers without considering topologies. Intel's oneCCL~\cite{oneccl} builds a tile/card/node communicator hierarchy, but triggers intra-GPU and inter-GPU communications  alltogether, leaving the communication order to hardware scheduling; moreover, oneCCL does not support asynchronous all-to-all, and hence cannot overlap all-to-all with computation. Recent work explores new communication substrates for GPU collectives, such as CXL-based memory pooling for GPU collectives~\cite{xu2026ccclnodespanninggpucollectives}, CXL-shared memory for MPI~\cite{10.1145/3712285.3759816}, and CXL-based tensor offloading for large model training~\cite{sc24_cxl}, but these approaches target different interconnect fabrics and do not address Aurora's intra-GPU/inter-GPU bandwidth hierarchy. We compare our approach against oneCCL's all-to-all in \S\ref{subsec:tapa-microbench}.

%% file: text/method.tex
\section{Design}
\label{sec:aura}
We present the design of \name in this section. 

\subsection{Overview}
\label{subsec:overview}

\begin{figure}[!t]
\centering
\includegraphics[width=1.01\linewidth]{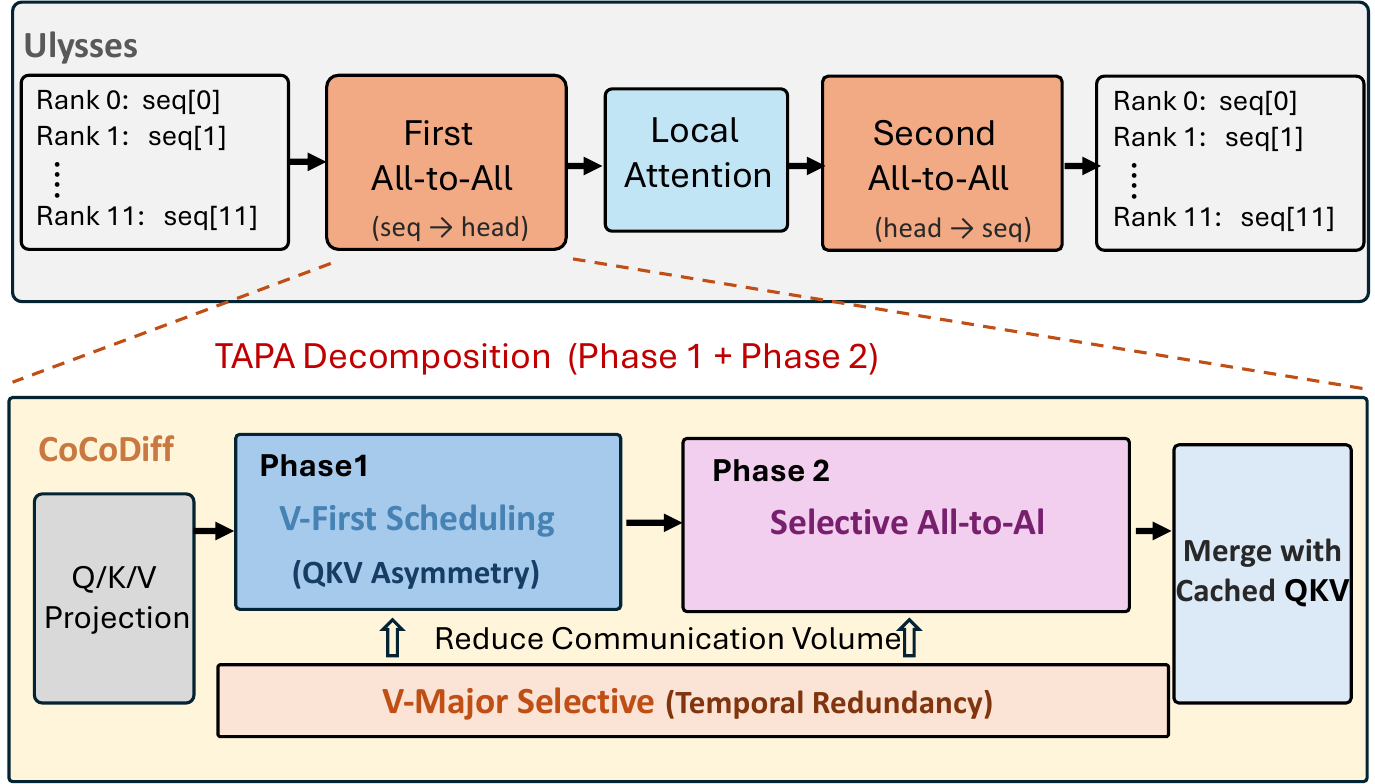}
\caption{\name overview.}
\label{fig:aura-architecture}
\vspace{-15pt}
\end{figure}

\name has three components, depicted in Figure~\ref{fig:aura-architecture}. The first component, ``TAPA decomposition'', decomposes each all-to-all communication into two phases where Phase 1 happens within each GPU using higher bandwidth and Phase 2 happens across GPUs using lower bandwidth. In each phase, multiple communications happen in parallel either within GPUs or across GPUs, leveraging aggregated communication bandwidth. The second phase especially benefits from the decomposition, because it allows two 6-rank all-to-all communications to occur in parallel, fully aligned with the interconnect topology and hence fully leveraging aggregated bandwidth across GPUs in a node.

The second component, V-First schedule, is designed for the first all-to-all communication in a layer. Instead of waiting for the projections for the three tensors ($Q$, $K$, and $V$) to be done altogether and then launching the first all-to-all for the three tensors, V-First launches the all-to-all for $V$ once the projection of $V$ is done. Because of the computation asymmetry across the three tensors (i.e, the lack of $QK$ normalization and RoPE for $V$), $V$ is able to finish the projection earlier and launch Phase 1 of its all-to-all. As a result, Phase 1 overlaps with the $QK$ normalization and RoPE.    

The third component, V-major schedule, is also designed for the first all-to-all communication in a layer. This component is very helpful to reduce the communication volume in the second phase of all-to-all, which happens in slower interconnect. This component caches 
projections in the prior time steps. In a time step, this component will compare the new projections against the cached projections, and identifies the active $q$, $k$, and $v$ vectors and communicates only those vectors instead of entire $Q$, $K$, and $V$.



In the rest of this section, we use a running example to make the discussion more concrete. In this example, we use  Qwen-Image (20B parameters and  60 layers) at the resolution $1536\times1536$ with batch size 1 on a single Aurora node (6 GPUs with 12 tiles in total), and we launch 12 ranks (i.e., one rank per tile). All timing numbers refer to this setup unless otherwise noted.



\subsection{Tile-Aware Parallel All-to-All}
\label{subsec:TAPA-decomposition}

\begin{figure}[!t]
\centering
\includegraphics[width=\linewidth]{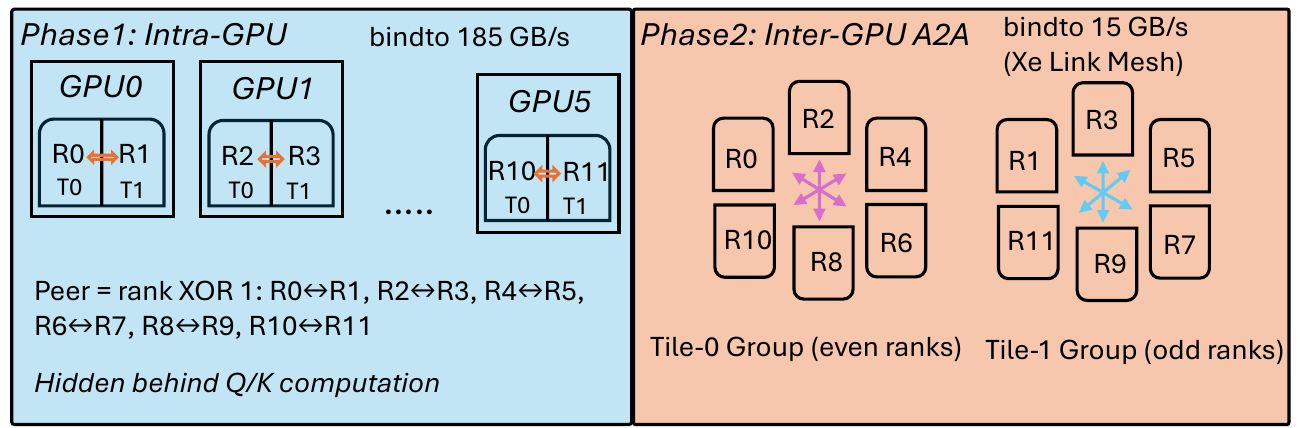}
\caption{Tile-aware parallel all-to-all.}
\label{fig:TAPA-phases}
\vspace{-15pt}
\end{figure}

Figure~\ref{fig:TAPA-phases} illustrates tile-aware parallel all-to-all (TAPA). In Phase~1, each GPU performs an intra-GPU tile-to-tile exchange. All six pairs of tiles perform internal token exchange in parallel, leveraging high bandwidth (185\,GB/s) of the internal links in GPUs. In our running example, this phase takes 0.3 ms to exchange messages within individual GPUs, and each message is 14\,MB. After Phase 1, each tile holds a copy of all tokens distributed to the GPU where the tile resides. Hence, when Phase 2 starts, each tile just needs to exchange tokens with tiles on other GPUs.

Phase 1 is based on point-to-point (p2p) communications. With the support of Intel GPU architecture and system software, those communications can occur asynchronously, allowing the overlap of those communications with computation (see V-First scheduling~\S\ref{subsec:vfirst}). In contrast, the original all-to-all communication cannot run asynchronously, because of the limitation of Intel system software. Hence, through the communication decomposition, TAPA provides performance optimization opportunities. 


Phase~2 performs inter-GPU communication as two parallel 6-rank all-to-alls, one per Xe Link ring. 
The two ring-based collectives execute in parallel with no cross-ring traffic, maximizing aggregated communication bandwidth.


The two-phase collective communication uses the above systematic plan to coordinate communication traffic within a node. Hence, the two-phase avoids any multi-hop routing from which a flat collective may suffer across the two rings.

\textbf{Some implementation details.} The asynchronous p2p communication in Phase 1 is implemented by a background hardware-thread per tile. In addition to the p2p communication itself, the computation thread offloads p2p initialization and tensor preparation (including reshaping, transposition, and buffer allocation) to the background thread to remove such overhead from the critical path.

\begin{figure}[!t]
\centering
\includegraphics[width=\linewidth]{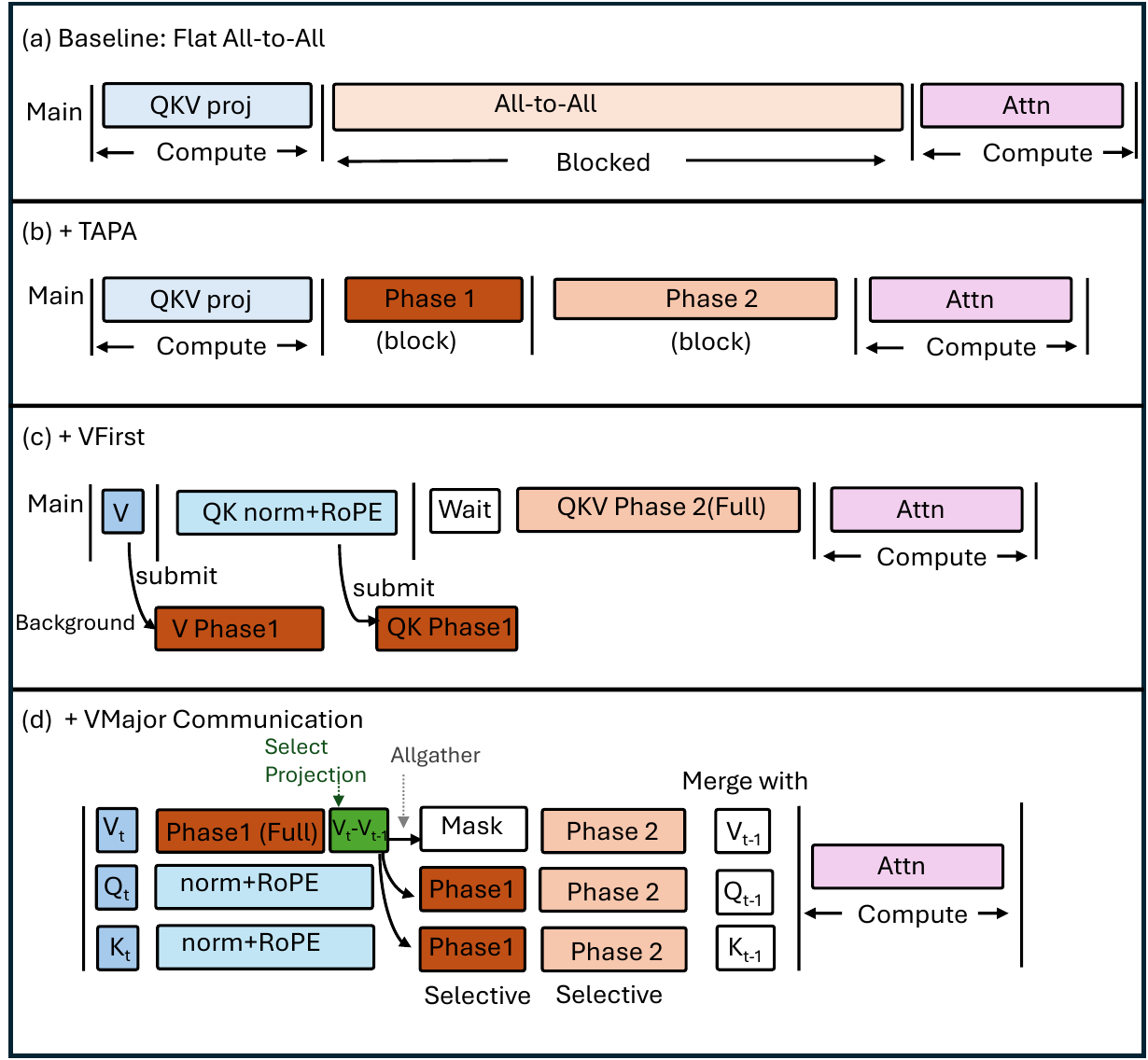}
\caption{Execution timeline comparison. (a) Baseline:  flat all-to-all in the critical path. (b) TAPA without communication offloading. (c) V-first scheduling with async offloading, hiding Phase~1 behind Q/K norm and RoPE. (d) V-Major selective communication: reducing the communication volume in QK's Phase~1 and Phase~2.}
\label{fig:timeline-comparison}
\end{figure}

\subsection{V-First Scheduling}
\label{subsec:vfirst}
V-First scheduling exploits the QKV processing asymmetry  (\S\ref{subsec:qkv-asymmetry}). In our running example, $V$'s projection takes 0.1\,ms, while $Q$ and $K$ each take 0.5\,ms because they additionally pass through RMSNorm and RoPE, leaving a 0.9\,ms gap between $V$ and $QK$.



With V-First scheduling, 
\textit{V}'s projection is first computed, immediately followed by \textit{V}'s Phase~1 communication occurring in the background; \textit{Q} and \textit{K} processing are performed at the same time with \textit{V}'s Phase~1 communication. While \textit{V}'s Phase~1 executes in the background, 
the computation threads perform \textit{Q}'s projection, normalization, and RoPE (0.5\,ms), and then \textit{K}'s projection, normalization, and RoPE (another 0.5\,ms). By the time \textit{Q} and \textit{K} are ready, \textit{V}'s Phase~1 has already completed---the 1.0\,ms of \textit{Q/K} processing fully overlaps with \textit{V}'s Phase~1 communication.

More formally, let $T(V^{p1})$ denote $V$'s Phase~1 latency and $T(Q,K)$ denote the total $Q$/$K$ processing time (normalization + RoPE). 
Phase~1 is completely hidden when the following condition is met. 
\begin{equation}
\label{eq:tqk_and_tv}
T(Q,K) > T(V^{p1})    
\end{equation}
In our running example, $T(Q,K) = 1.0$\,ms and $T(V^{p1}) = 0.3$\,ms, which satisfies this condition.  
Equation~\ref{eq:tqk_and_tv} holds across all DiT models we evaluated (i.e., FLUX, SD3.5, and Qwen-Image).  

However, there is a risk of violating the inequality in Equation~\ref{eq:tqk_and_tv} when the batch size is large. In particular, $V$'s projection time scales linearly with the batch size and can grow faster than $T(Q,K)$, As a result, there is less time gap for V-First to tap; $T(V^{p1})$ may not be completely overlapped with $T(Q,K)$ and hence exposed to the critical path. Even so, $T(V^{p1})$ can still be partially hidden.

\subsection{V-Major Selective Communication}
\label{subsec:vmajor}

V-major reduces the amount of data exposed to inter-GPU and intra-GPU communications. The key insight is that diffusion inference exhibits significant temporal redundancy across denoising steps~\cite{ma2024deepcache, chen2024deltadit, ma2024learningtocache, liu2025teacache}: adjacent timesteps produce similar intermediate representations (i.e., $q$, $k$, and $v$ vectors), and this redundancy can be exploited to communicate only those $q$, $k$, and $v$ (i.e., the token projections) whose corresponding tokens have changed significantly, while reusing cached ones for the remainder.


\textbf{Active projection selection.} To identify which tokens have changed prominently between consecutive time steps, we need an indicator per token. Any of the three projections for a given token could serve this role, since they are all derived from the hidden state of the same token.


For each token, we use one of its projections (particularly the value vector $v$) against the cached $v$ from previous time steps to build an active-token mask. In particular, at a time step $t$, after Phase 1 of $V$ is done, each tile in a GPU holds the projections for all tokens assigned to that GPU, and one of the tiles then compares all the $v$ it has collected against the cached ones from the prior time steps. The comparison is based on a scalar metric, which is the $L_1$ norm of the distance of two value vectors. A larger metric value indicates big changes, and vice versa. Among all the vectors for comparison, we choose a fraction ($r$) of them with the smallest metric values to be cached, and those vectors are considered stable and are not communicated across GPUs. $r$ is called the cache ratio. The remaining $(1-r)$ of them will be used for communication. $r \in [0,1]$ is a controllable hyperparameter, and dynamically changed at runtime (see ``Time-varying cache schedule'' for details).   


After each GPU selects its active projections, the indices of active projections are then shared across all GPUs via an all-gather collective to ensure consistent projection selection across all GPU tiles. 
This collective is cheap as it communicates only a small integer-index list. 

\textbf{Why using $v$ for  projection selection?}  We select the active projections based on $v$ (not $q$ or $k$).  This method is based on the assumption that the variance of $v$ across time steps is an indication of the variance of $k$ and $q$ across the same time steps. This method is effective for projection selection, because $q$, $k$, and $v$ are linear projections of the same hidden state of the token, hence their value variances across time steps follow the same trend.

Furthermore, we use $v$, instead of $q$ and $k$, because the projection $v$ is finished earlier than the other two projections, hence allowing us to start the projection  selection earlier and overlap it with ``norm+RoPE''.

We do not directly use tokens to calculate the changes across time steps, because using $v$ after Phase 1 allows us to overlap the projection selection with ``norm+RoPE'', shown in Figure~\ref{fig:timeline-comparison}.d, hence hiding the selection overhead.



\textbf{Selective communication in TAPA.} 
We discuss the communication in TAPA with V-major in place as follows.

\textit{Phase 1 (intra-GPU tile-to-tile):} $V$'s Phase 1 communication has to be done at full size, as the per-GPU projection selection requires the L1 norm of the complete exchange output in Phase 1. For $Q$ and $K$, only the active projections are extracted before the tile-to-tile exchange, reducing the Phase 1 data volume to $(1-r)$ of its original size. See Figure~\ref{fig:timeline-comparison}.d. 

\textit{Phase 2 (inter-GPU all-to-all):} All the three projections undergo inter-GPU all-to-all, but only the active subset is communicated. This is where the largest savings occur: the message size drops by a factor of $1/r$, directly reducing traffic on the bandwidth-limited inter-GPU links.


\textit{Post-Phase 2 (Receiver-side data reconstruction):} Each GPU tile caches active projections from the time steps. Those projections are in the head-parallel layout (\S\ref{subsec:seq-parallel}). After receiving those active projections in Phase 2, each tile goes through a reconstruction process to reconstruct $Q_t$, $K_t$, and $V_t$ for the time step $t$, which include active projections in $t$ and cached projections.

The reconstruction is an indexed scatter: at the time step $t$, each tile uses the all-gathered indices of active projections to overwrite into $Q_{t-1}$, $K_{t-1}$, and $V_{t-1}$. As a result, the inactive projections keep their cached values unchanged. The above reconstruction produces a complete  $Q_{t}$, $K_{t}$, and $V_{t}$, assembled with minimal data movement and ready for attention. 

\textbf{Time-varying cache scheduling.}
We exploit the cache management with three complementary mechanisms.
\emph{(i) Warmup.} The selective path requires a cached baseline from the previous timestep, so the first few denoising steps run full TAPA to populate the cache; selective communication begins from step $\text{warmup}+1$ onward.
\emph{(ii) Time-varying cache ratio.} The cache ratio $r$ trades off communication reduction against inference accuracy:  a higher $r$ skips more projections for greater speedup but risks degrading fine details, while lower $r$ preserves quality at higher cost. A fixed $r$ thus locks the system to a single point on this trade-off curve. DiT, however, exhibits stage-dependent dynamics: early steps require large global updates, whereas later steps involve smaller, localized refinements, reducing the number of projections that must be communicated over time. From the warmup boundary onward, $r$ follows a \textit{linear} schedule from $0$ to $1$ across time steps, communicating most projections early, and caching aggressively late. 
\emph{(iii) Periodic full synchronization.} Because each selective step updates only the active projections, 
the remaining entries retain values from earlier time steps; as denoising progresses, unrefreshed cache entries may carry data from $t{-}1$, $t{-}2$, or even older steps, causing approximation error to accumulate across layers. At a configurable interval (10 by default), we therefore force a full-size update to flush these stale entries and reset the cache to the exact values at the current timestep.

\textbf{Memory overhead analysis.} Managing cached tensors efficiently is critical for inference workloads on memory-constrained devices~\cite{hpca25:dlrm, hpca21:ren}. Each transformer layer caches four tensors, namely $V^{p1}$ (from the Phase 1 exchange) and $Q^{p2}$, $K^{p2}$, and $V^{p2}$ (from the Phase 2 all-to-all).  

Two properties follow. First, total overhead grows linearly with the batch size, since each of the four tensors scales with the number of latents processed per tile. Second, per-tile overhead shrinks as the total number of tiles increases: doubling the resources halves each tile's share of the sequence and heads, so all the four cached tensors shrink by the same factor. For example, Qwen-Image at $1536\times1536$ with $B{=}4$ on single node (12 tiles) requires 4{,}320\,MB of cache per-tile, roughly 6.6\% of the 64\,GB tile memory capacity. Scaling to 8 nodes (96 tiles)  with $B{=}32$ increases the batch size by $8\times$ while distributing the cache across $8\times$ more tiles, keeping the per-tile overhead unchanged.



\subsection{Multi-Node Extension}
\label{subsec:multi-node}
\name's two-phase structure can be extended to multiple nodes without modification: Phase~1 remains a strictly intra-GPU exchange, and Phase~2 is a collective that routes across Xe Link within node and across nodes. 

We do not split Phase 2 further into intra-node and inter-node phases on Aurora because of the following reason. 
On Aurora, intra-node Xe Link bandwidth (15\,GB/s) and inter-node Slingshot bandwidth (25\,GB/s) are comparable. The bandwidth difference (1.67$\times$) between the two links are much smaller than that (12$\times$) between intra-GPU and inter-GPU within a node. 
Moreover, in the context of DiT inference, splitting Phase 2 does not introduce any opportunity to overlap with computation, but adds synchronization overhead.




\algnewcommand{\LineComment}[1]{\State \(\triangleright\) \textit{#1}}
\algnewcommand{\SideComment}[1]{\hfill\(\triangleright\) \textit{#1}}

%% file: text/evaluation.tex
\section{Evaluation}
\label{sec:evaluation}

We evaluate \name on Intel Aurora, demonstrating significant communication speedup and end-to-end performance improvement across multiple DiT models and configurations. \name is implemented as an extension to xDiT~\cite{fang2024xdit} and the Diffusers~\cite{von-platen-etal-2022-diffusers} library, contributing approximately 20K lines of code. Our code is publicly available at \url{https://anonymous.4open.science/r/xDit-intel-2CB8/} and \url{https://anonymous.4open.science/r/sp-aurora-92F3/}.

\subsection{Experimental Setup}
\label{subsec:setup}

\textbf{Hardware.} 
On Aurora, each node is equipped with six Intel Data Center GPU Max 1550 (Ponte Vecchio) accelerators, each comprising two compute tiles with 64,GB HBM2e per tile. In total, a node provides 12 addressable compute tiles interconnected via Xe Links (15\,GB/s per direction between GPUs) and high-bandwidth intra-GPU fabric (185\,GB/s between tiles). 
For single-node experiments, we use all 12 tiles and for multi-node experiments, we scale to 2, 4, and 8 nodes connected via eight HPE Slingshot-11 NICs per node (25\,GB/s unidirectional per NIC).

\textbf{Software.} We use PyTorch 2.10.0 with Intel Extension (IPEX) and Intel oneCCL 2021.17 for collective communication. 
Unless otherwise specified, we use a Ulysses degree of 12 on a single node, 24 on two nodes, and 48 on four and eight nodes, with 50 denoising steps, 5 warmup iterations, and 10-step periodic synchronization. All results are averaged over three runs. 

\textbf{Models and configurations.} We evaluate four DiT models: FLUX.1-dev~\cite{flux2024} (12B parameters and 57 blocks), Qwen-Image~\cite{wu2025qwenimagetechnicalreport} (20B parameters and 60 blocks), Stable Diffusion 3.5 (SD3.5)~\cite{esser2024sd3} (8B parameters and 38 blocks), and FLUX.2-dev~\cite{flux-2-2025} (32B parameters and 56 blocks). We vary the image resolution ($768\times768$, $1536\times1536$, and $2304\times2304$) and batch size (1, 2, 4, 8, 16, and 32).

\textbf{Baselines.} We compare against three configurations: (1) \textit{Flat}, standard Ulysses all-to-all; (2) \textit{TAPA}, tile-aware parallel all-to-all; (3) \textit{\name}, TAPA augmented with V-First scheduling and V-Major selective communication.

\textbf{Metrics.} We measure \name's end-to-end speedup and inpainting quality via Peak Signal-to-Noise Ratio (PSNR)~\cite{hore2010image} and Structural Similarity Index (SSIM)~\cite{wang2004image}, along with the resulting speedup–quality trade-off.

\textbf{Dataset.} We evaluate our approach on a scientific inpainting task using synchrotron X-ray microtomography data~\cite{mlr} of mouse brain tissue acquired at the Advanced Photon Source (APS) at Argonne National Laboratory (ANL).

\subsection{End-to-End Speedups}
\label{subsec:speedups}

\begin{figure*}[t]
\centering
\subfloat[Single-node speedups (batch size $B \in\{1,2,4\}$).]{
\includegraphics[width=\textwidth]{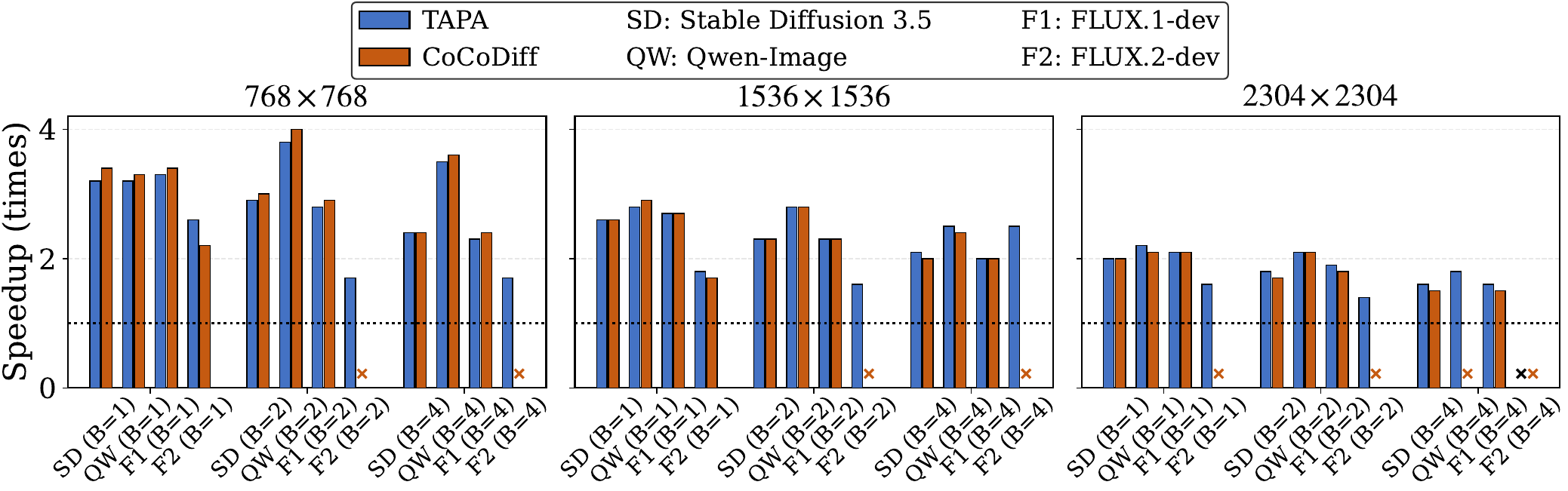}
\label{subfig:single}
}
\\
\subfloat[Multi-node speedups (node count $N \in \{2,4,8\}$). Each model is evaluated with the largest batch size that fits within GPU memory.]{
\includegraphics[width=.96\textwidth]{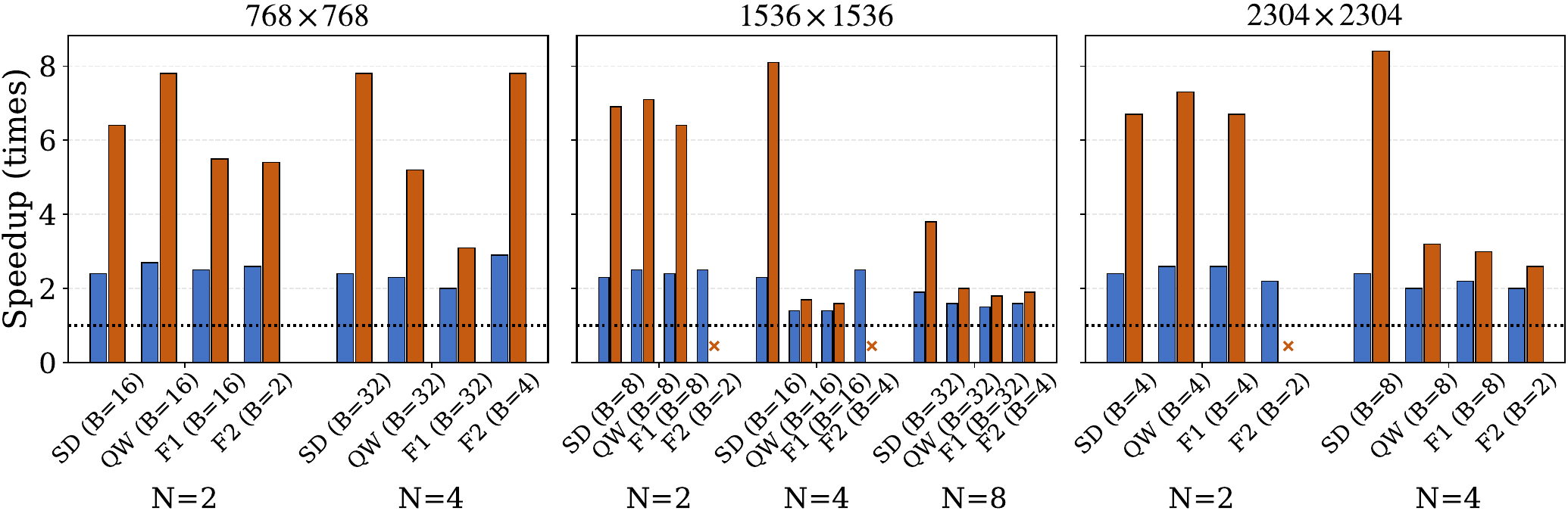}
\label{subfig:multi}
\vspace{-15pt}
}

\caption{End-to-end speedups of TAPA (blue) and \name (orange) over the Flat baseline across three image resolutions and four DiT models. ``$\times$'' denotes OOM due to model memory limits, while ``\textcolor{orange}{$\times$}'' denotes OOM caused by \name's cache mechanism. The absence of 8-node bars at $768\times768$ and $2304\times2304$ reflects a model-imposed layout constraint rather than a limitation of our communication strategy.}
\label{fig:speedups}
\vspace{-15pt}
\end{figure*}

We evaluate \name against Flat (baseline) and TAPA across three image resolutions ($768\times768$, $1536\times1536$, and $2304\times2304$) and four DiT models (SD3.5, Qwen-Image, FLUX.1-dev, and FLUX.2-dev). Figure~\ref{fig:speedups} presents the results as side-by-side bar charts. 
Across all configurations, \name delivers an average speedup of $3.6\times$ over Flat, compared to $2.3\times$ for TAPA, and reaches up to $8.4\times$ on SD3.5 at $2304\times2304$ with 4 nodes.

\textbf{Single-node speedups (Figure~\ref{subfig:single}).} 
\name and TAPA achieve comparable speedups over Flat ($2.5\times$ vs. $2.3\times$ on average). The limited advantage of \name over TAPA stems from the overhead of V-Major selective caching, including active projection selection, cache management, and an additional all-gather operation, which could offset its communication savings at this small scale.


\textbf{Multi-node speedups (Figure~\ref{subfig:multi}).}
\name achieves substantially higher speedups than TAPA over Flat ($5.1\times$ vs. $2.2\times$ on average), as inter-node all-to-all latency dominates at scale, making V-Major selective communication particularly effective.

Speedup, however, does not increase monotonically with node count. Ulysses parallelism is bounded by the model’s attention head count (one head per rank); once saturated, additional ranks no longer contribute to Ulysses all-to-all and are instead mapped to the Ring dimension. Because \name optimizes only Ulysses communication while leaving K/V circulation in the Ring dimension unchanged, further scaling increases the fraction of unoptimized communication. As a result, the marginal benefit of adding nodes diminishes beyond the Ulysses limit and can even reverse. 

This effect is evident in Figure~\ref{subfig:multi}. Qwen-Image (24 heads) at $1536\times1536$ saturates at 2 nodes and drops sharply at 4 nodes (from $7.1\times$ to $1.7\times$). In contrast, SD3.5 (48 heads) at the same resolution saturates at 4 nodes and drops at 8 nodes (from $8.4\times$ to $3.8\times$).


\textbf{Out-of-memory configurations.}
We observe two distinct out-of-memory (OOM) scenarios. First, the model alone can exhaust GPU memory even under the Flat baseline, as indicated by ``$\times$'' in Figure~\ref{fig:speedups}. Second, for FLUX.2-dev, the 32B model weights leave limited memory headroom, causing \name's caching mechanism to trigger OOM at larger batch sizes, marked by ``\textcolor{orange}{$\times$}''.

\textbf{Absent entries in multi-node configurations.}
The absence of 8-node bars at $768\times768$ and $2304\times2304$ arises from a divisibility constraint in DiT sequence parallelism. An $H\times H$ image is encoded into a $H/8\times H/8$ latent grid, and each $2\times2$ block is grouped into a single token, yielding a $H/16\times H/16$ token grid~\cite{esser2024sd3}. The sequence parallelism partitions tokens evenly across $P$ tiles along the sequence dimension, which requires $(H/16) \bmod P = 0$ to ensure balanced workloads and avoid cross-tile dependency violations. 

Consequently, at 8 nodes (96 tiles), this condition is not satisfied for $768\times768$ and $2304\times2304$, which produce token grids of $48\times48$ and $144\times144$, respectively; neither can be evenly divided across 96 tiles. In contrast, $1536\times1536$ yields a $96\times96$ token grid, which aligns with the sequence parallelism and can be evenly distributed. 

This limitation is inherent to the model’s tokenization and parallelization scheme, rather than a restriction of our communication strategy.

\begin{figure}[th]
\centering
\includegraphics[width=\columnwidth]{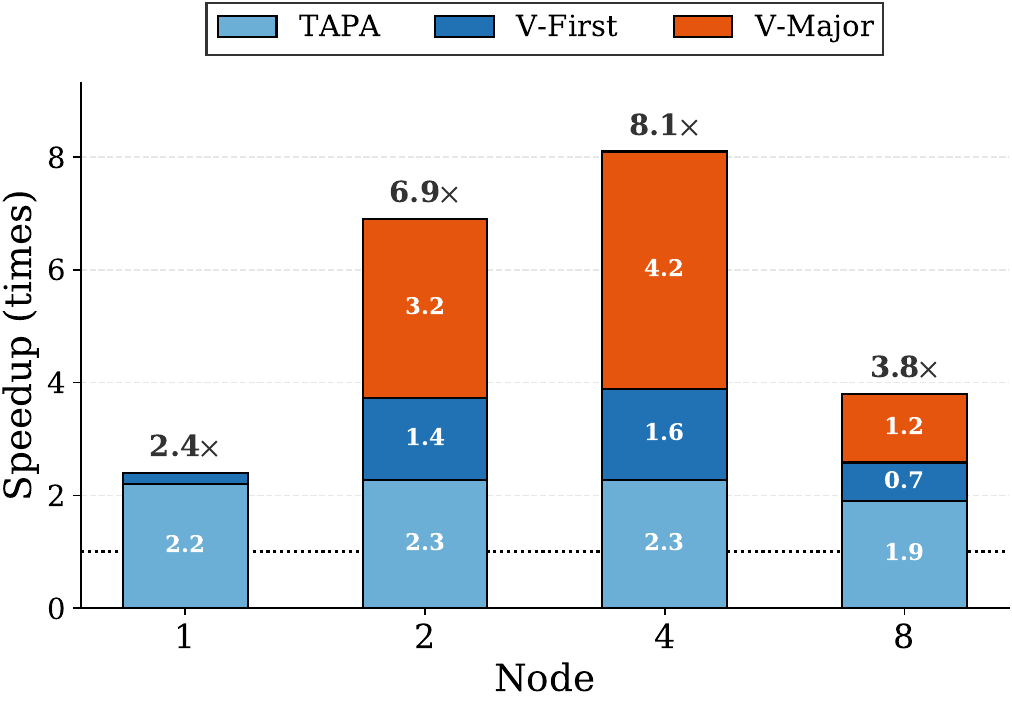}
\caption{Speedup breakdown of \name on SD3.5 at $1536\times1536$. Each stacked bar decomposes the total speedup into contributions from TAPA, V-First, and V-Major across 1-, 2-, 4-, and 8-node configurations.}
\vspace{-10pt}
\label{fig:ablation}
\end{figure}

\subsection{Ablation Studies}
\label{subsec:ablation}


We evaluate the contribution of each \name component—TAPA, V-First scheduling, and V-Major selective communication—by enabling them incrementally on SD3.5 at a resolution of $1536\times1536$ across 1, 2, 4, and 8 nodes. Figure~\ref{fig:ablation} decomposes \name’s end-to-end speedup into these components at each scale. We make three observations.

First, TAPA provides a near-constant baseline speedup (averaging 2.2$\times$) across all configurations by decomposing the monolithic all-to-all into two phases aligned with the hardware hierarchy. 
Second, V-First scheduling and V-Major selective communication yield negligible gains on a single node but become dominant contributors at scale, achieving average speedups of $1.3\times$ and $2.9\times$ across 2-, 4-, and 8-node configurations. 
Third, from 4 to 8 nodes, these gains diminish. Ulysses parallelism is bounded by the model’s attention head count (48 heads in SD3.5), and this limit is reached at 4 nodes. Beyond this point, additional tiles at 8 nodes no longer expand Ulysses parallelism and are instead mapped to the Ring dimension. 
Because \name optimizes only Ulysses communication while leaving the Ring dimension unoptimized, further scaling increases the share of unoptimized communication, thereby reducing the overall gains.

\subsection{Quality Evaluation of Restored Medical Image}
\label{subsec:ImgQuality}

\begin{figure}[t]
\centering
\includegraphics[width=\linewidth]{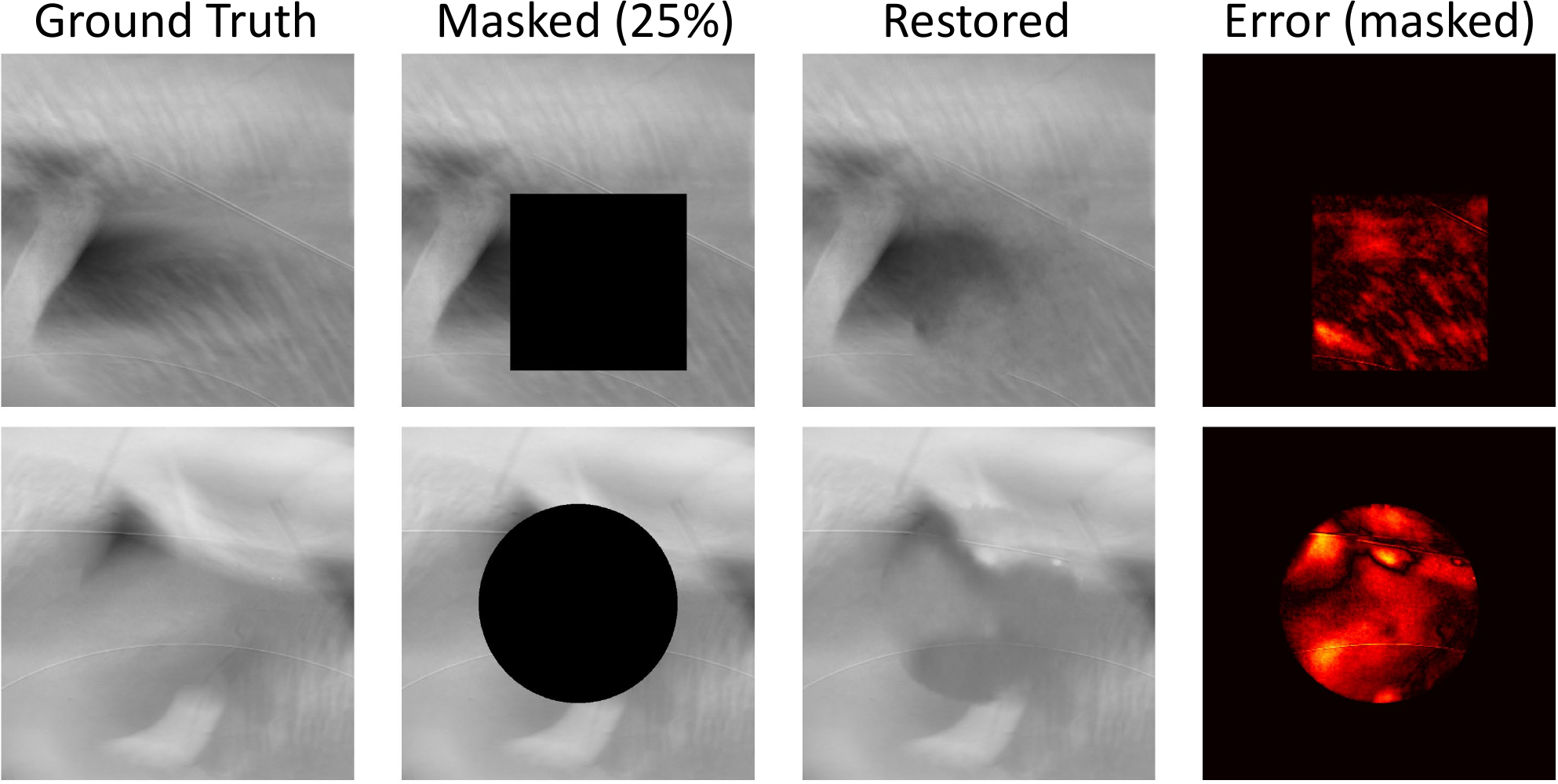}
\caption{Inpainting results of FLUX.1-dev with \name on a mouse brain tissue slice. From left to right: original images ($768\times768$), masked inputs (block and center masks with 25\% coverage), restored outputs, and pixel-wise error maps (brighter colors indicate larger errors).}
\label{fig:inpainted}
\end{figure}

Table~\ref{tab:inpaint-quality} reports inpainting quality metrics---PSNR in decibels (dB; higher is better), SSIM (higher is better), and mean squared error (MSE)---for FLUX.1-dev, SD3.5, and Qwen-Image\footnote{FLUX.2-dev is excluded because the Diffusers library~\cite{von-platen-etal-2022-diffusers} does not yet provide an inpainting pipeline for this model.}. Results are evaluated on X-ray microtomography data~\cite{mlr} of mouse brain tissue using block and center masks, as visualized in Figure~\ref{fig:inpainted}. 

Across all models, mask types, and metrics, Flat, TAPA, and \name produce comparable image quality. Differences remain small within 2\,dB for PSNR, 0.03 for SSIM, and $2\times10^{-3}$ for MSE, indicating that neither hierarchical communication restructuring nor selective projection  caching introduces non-trivial degradation relative to Flat (serving as the reference output). Overall, these results show that \name preserves image quality within the noise floor of the generative process. 
This holds even for inpainting, where comparison against Flat provides a stringent evaluation, as pixel-wise metrics such as PSNR, SSIM, and MSE directly capture any deviations introduced by communication optimizations relative to full, unapproximated execution.

\begin{table}[t]
\centering
\caption{Inpainting quality metrics evaluated under block and center mask configurations.}
\label{tab:inpaint-quality}
\setlength{\tabcolsep}{3pt}
\renewcommand{\arraystretch}{1.5}
\resizebox{\columnwidth}{!}{
\begin{tabular}{l|ccc|ccc|ccc}
\hline
\textbf{Metric} 
& \multicolumn{3}{c|}{\textbf{SD3.5}} 
& \multicolumn{3}{c|}{\textbf{FLUX.1-dev}} 
& \multicolumn{3}{c}{\textbf{Qwen-Image}} \\
& Flat & TAPA & \name & Flat & TAPA & \name & Flat & TAPA & \name \\
\hline

\multicolumn{10}{c}{\textbf{Block Masked Region}} \\
\hline
PSNR (dB)
& 23.05 & \textbf{23.11} & 22.77
& 26.20 & \textbf{27.07} & 25.05
& \textbf{29.97} & 29.63 & 29.17  \\

SSIM
& 0.8036 & \textbf{0.8083} & 0.8042 
& 0.8879 & \textbf{0.8928} & 0.8816
& \textbf{0.9256} & 0.9248 & 0.9229   \\

MSE
& 0.0065 & \textbf{0.0051} & 0.0081 
& 0.0026 & \textbf{0.0020} & 0.0032
& \textbf{0.0011} & 0.0012 & 0.0012  \\

\hline

\multicolumn{10}{c}{\textbf{Center Masked Region}} \\
\hline
PSNR (dB)
& \textbf{23.65} & 22.15 & 21.95
& 24.89 & \textbf{24.91} & 24.22 
& \textbf{29.46} & 28.91 & 28.56   \\

SSIM
& \textbf{0.8390} & 0.8110 & 0.8075
& 0.8717 & \textbf{0.8838} & 0.8672
& \textbf{0.9236} & 0.9232 & 0.9233  \\

MSE
& \textbf{0.0046} & 0.0065 & 0.0066
& 0.0039 & \textbf{0.0035} & 0.0042
& \textbf{0.0012} & 0.0014 & 0.0015   \\

\hline
\end{tabular}
}
\end{table}

\subsection{Trade-off Between Image Quality and Speedup}
\label{subsec:QualityvsSpeedups}

\begin{figure}[t]
\centering
\subfloat[Trade-off between PSNR and Speedup]{
\includegraphics[width=\columnwidth]{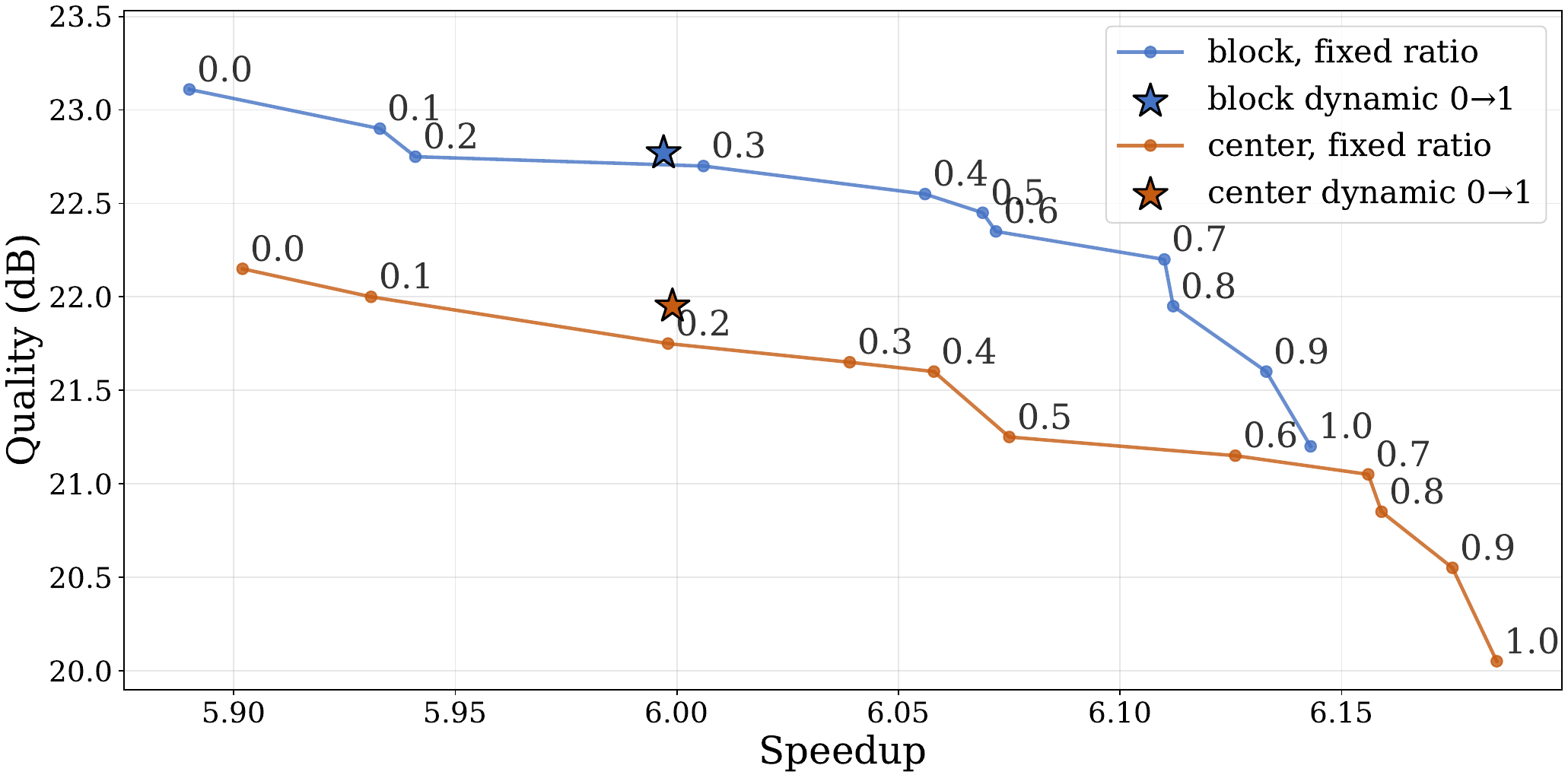}
\label{subfig:psnr-vs-speedup}
}\\
\subfloat[Trade-off between SSIM and Speedup]{
\includegraphics[width=\columnwidth]{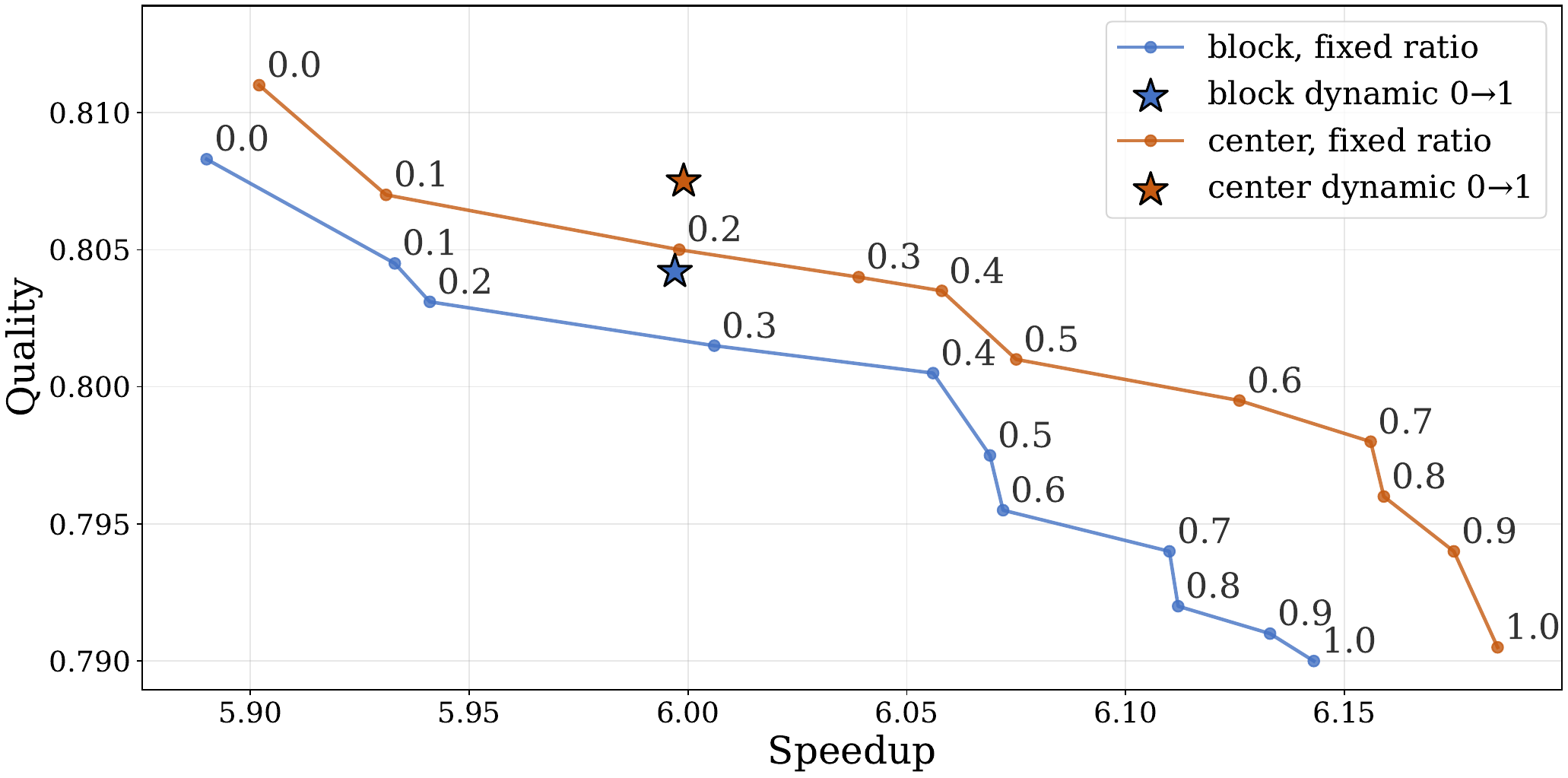}
\label{subfig:ssim-vs-speedup}
}
\caption{Inpainting quality vs. speedup for SD3.5 at $768\times768$. The time-varying cache ratio ($\bigstar$) consistently dominates the fixed-ratio baselines ($\bullet$): for a given speedup, it achieves higher inpainting quality, and for a given quality level, it attains a greater speedup.}
\vspace{-15pt}
\label{fig:qualityvsspeedups}
\end{figure}

Figure~\ref{fig:qualityvsspeedups} plots inpainting quality vs. speedup for SD3.5 at $768\times 768$, which compares fixed cache ratios (marked in $\bullet$) with time-varying cache ratios (marked in $\bigstar$). Quality is measured on the restored masked regions (block and center) using PSNR and SSIM. We make two observations. 
First, as expected, higher cache ratios yield greater speedups but lower inpainting quality, reflecting increased reuse of cached representations. 
Second, the time-varying cache ratio consistently dominates the fixed-ratio baselines: for a given speedup, it achieves higher inpainting quality, and for a given quality level, it attains a greater speedup. This improvement arises from its ability to adapt communication to evolving denoising dynamics: it preserves accuracy during early stages of DiT, when updates are large, and reduces communication aggressively in later stages as representations stabilize.


\subsection{TAPA vs. oneCCL for All-to-All}
\label{subsec:tapa-microbench}
\begin{figure}[!t]
\centering
\includegraphics[width=\columnwidth]{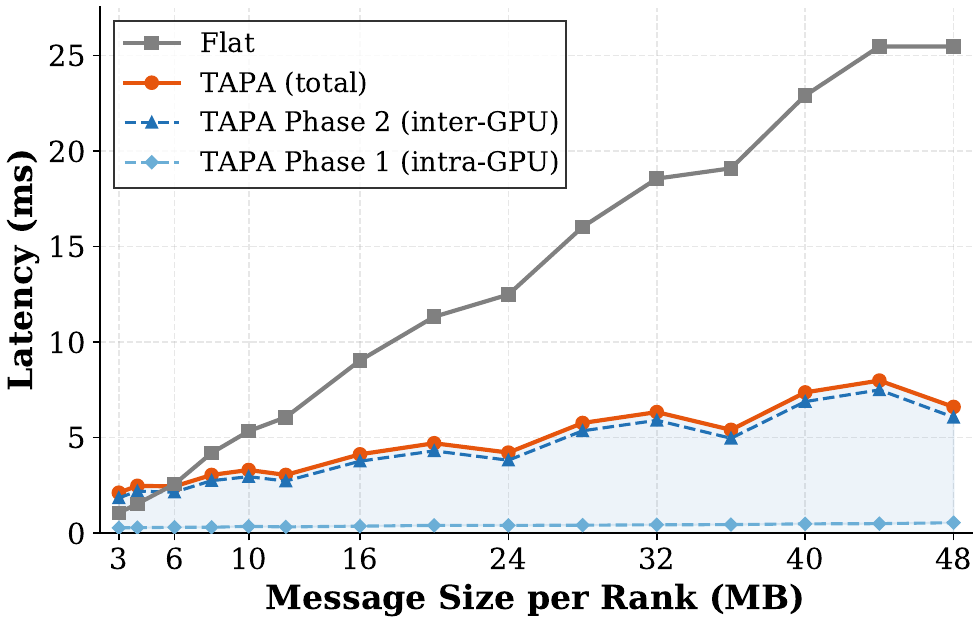}
\caption{ All-to-all communication latency on a single Aurora node (12 tiles).}
\label{fig:alltoall-latency}
\vspace{-15pt}
\end{figure}

We compare TAPA with Intel oneCCL with different message sizes. In this comparison, we focus on the communication only. 
We measure communication latency on a single node (12 tiles), and change the message size per rank from 3 to 48\,MB, the per-rank message size range commonly encountered in Ulysses sequence-parallel DiT inference. Figure~\ref{fig:alltoall-latency} reports the latency for oneCCL (grey), TAPA total (orange), and the two TAPA phases.  

The curve for Phase~1 (light blue) stays at 0.3--0.5\,ms across all message sizes, confirming that intra-GPU fabric provides high bandwidth (185\,GB/s) to handle tile-to-tile exchange with negligible overhead. 
The curve for Phase~2 (deep blue) dominates TAPA’s total latency and accounts for the gap between Phase~2 and TAPA (total). Because Phase~2 runs the two rings in parallel rather than one 12-tile collective, it reaches only 6.0\,ms at 48\,MB, while oneCCL’s 12-tile all-to-all reaches 25.5\,ms. In this case, TAPA reduces the latency by $3.9\times$. 

The advantage of TAPA becomes larger with larger message sizes: at 6\,MB, Flat and TAPA (total) have comparable performance, but at 24\,MB, TAPA outperforms by $3.0\times$. When the message size is 48\,MB, TAPA outperforms Flat by $3.9\times$.


%% file: text/related_work.tex
\section{Related Work}
\label{sec:related}

\textbf{Distributed diffusion transformer inference.} Several systems target distributed inference specifically for diffusion models. DistriFusion~\cite{li2024distrifusion} introduces patch parallelism, partitioning spatial patches across GPUs and reusing stale interaction from the previous denoising step to avoid synchronous communication. PipeFusion~\cite{wang2024pipefusion} applies pipeline parallelism across transformer layers, exploiting temporal redundancy to overlap computation of the current step with stale activations from prior steps. AsyncDiff~\cite{chen2024asyncdiff} parallelizes denoising by running consecutive steps on different GPUs asynchronously. xDiT~\cite{fang2024xdit} provides a hybrid engine combining Ulysses, Ring, PipeFusion, and CFG parallelism. More recently, ScaleFusion~\cite{chen2025scalefusion} targets video DiTs by scheduling intra-layer and inter-layer communication to hide cross-machine overhead, achieving 3.6$\times$ strong scaling on 32 A100s. StreamFusion~\cite{luo2026streamfusion} designs a topology-aware serving engine using NVSHMEM to unify Torus, Ulysses, and Ring communication on NVIDIA GPUs.

The above systems optimize parallelism strategies but treat the all-to-all as a black box. \name is complementary: it restructures and reduces the all-to-all itself by exploiting QKV asymmetry and temporal redundancy, and is the first distributed DiT inference engine optimized for Intel GPU clusters.

\textbf{Cache-Based Diffusion Inference Acceleration.} 
There are existing efforts that use cache-based methods to exploit temporal redundancy across denoising steps to skip redundant single-GPU computation~\cite{ ma2024learningtocache, liu2025cachequant, ma2025adacache}. Diffusion Transformers are especially amenable to such caching, and several DiT-specific techniques have emerged~\cite{chen2024deltadit, liu2025taylorseer, zhao2025speca}. DeepCache~\cite{ma2024deepcache} and $\Delta$-DiT~\cite{chen2024deltadit} cache intermediate features across steps, skipping computation for layers whose outputs change little. Learning-to-Cache~\cite{ma2024learningtocache} trains a policy to adaptively select which layers to cache at each step. TaylorSeer~\cite{liu2025taylorseer} goes beyond reuse by predicting future features via Taylor expansion, achieving nearly 5$\times$ speedup on FLUX. SpeCa~\cite{zhao2025speca} applies a speculative-decoding-style forecast-then-verify framework to DiTs. CacheQuant~\cite{liu2025cachequant} co-optimizes caching schedules with quantization via dynamic programming. AdaCache~\cite{ma2025adacache} makes caching content-adaptive, allocating more computation to high-motion video regions.

Memoization techniques share a similar principle of exploiting computation similarity to reduce redundant computation in iterative workloads~\cite{mlr, ics21:memoization}.
The above methods reduce compute on a single device. \name's V-Major targets a different bottleneck: communication across devices, using temporal redundancy to reduce communication volumes. 

\textbf{Sequence parallelism for long-context attention.}
Sequence parallelism distributes attention across devices along the sequence dimension. 
Ulysses uses all-to-all collectives to transform between sequence-parallel and head-parallel layouts (\S\ref{subsec:seq-parallel}), with DeepSpeed-Ulysses~\cite{jacobs2023deepspeed} and Megatron-LM~\cite{shoeybi2019megatron, korthikanti2023reducing} adopting similar schemes. Ring Attention~\cite{liu2024ring} circulates K/V blocks in a ring, overlapping communication with computation within each round. Striped Attention~\cite{brandon2023striped} and LightSeq~\cite{li2023lightseq} refine load balance and scheduling. USP~\cite{fang2024usp} and LoongTrain~\cite{gu2024loongtrain} apply Ulysses within nodes and Ring across nodes.

\name is orthogonal to the above efforts: it optimizes the all-to-all communications commonly employed in the sequence parallelism. 

%% file: text/conclusions.tex
\section{Conclusions}
\label{sec:conclusion}

We present \name, a distributed DiT inference engine that leverages application-level knowledge, specifically QKV processing asymmetry and temporal redundancy, into the collective communication layer. \name uses three synergic mechanisms: TAPA creates topology-aligned communications, V-First enables overlap between communication and computation, 
and V-Major reduces the remaining traffic by transmitting only active projections. Deployed on the Aurora supercomputer across four DiT models scaling to 96 GPU tiles, \name averages 3.6$\times$ speedup, peaking at $8.4\times$, while preserving image quality, enabling efficient large-scale DiT inference. The principle of co-designing communication with application semantics and hardware topology generalizes to any platform with hierarchical interconnects.